\documentclass[aip,preprint,showpacs,endfloats*,a4paper]{revtex4-1}
\usepackage{graphicx}
\usepackage{epsfig}
\usepackage{dcolumn}
\usepackage{epsfig}
\usepackage[T1]{fontenc}
\usepackage{currvita}
\usepackage{amsmath}
\usepackage{times,mathptm}
\begin{document}

\title{GaAs/GaP quantum dots: Ensemble of direct and indirect heterostructures with room temperature optical emission}
% Force line breaks with \\

\author{S. Dadgostar}
\affiliation{%
Department of Physics, Humboldt-Universit\"{a}t zu Berlin, Newton-Str. 15, D-12489 Berlin, Germany
%\\This line break forced% with \\
} %
%\email{dadgostar@physik.hu-berlin.de}

\author{J. Schmidtbauer}
\affiliation{%
Leibniz-Institut f\"{u}r Kristallz\"{u}chtung, Max-Born-Str. 2, D-12489 Berlin, Germany%\\This line break forced% with \\
}%
\author{T. Boeck}
\affiliation{%
Leibniz-Institut f\"{u}r Kristallz\"{u}chtung, Max-Born-Str. 2, D-12489 Berlin, Germany%\\This line break forced% with \\
}%

\author{A. Torres}
\affiliation{%
Departamento de Fisica de la Materia Condensada, E.T.S.I.I., 47011, Valladolid, Spain%\\This line break forced% with \\
}%

\author{O. Mart\'{i}nez}
\affiliation{%
Departamento de Fisica de la Materia Condensada, E.T.S.I.I., 47011, Valladolid, Spain%\\This line break forced% with \\
}%

\author{J. Jim\'{e}nez}
\affiliation{%
Departamento de Fisica de la Materia Condensada, E.T.S.I.I., 47011, Valladolid, Spain%\\This line break forced% with \\
}%

\author{J. W. Tomm}
\affiliation{%
Max-Born-Institut f\"{u}r Nichtlineare Optik und Kurzzeitspektroskopie,  Max-Born-Str. 2A, 12489 Berlin,
Germany%\\This line break forced% with \\
}%

\author{A. Mogilatenko}
\affiliation{%
Department of Physics, Humboldt-Universit\"{a}t zu Berlin, Newton-Str. 15, D-12489 Berlin, Germany
%\\This line break forced% with \\
}%

\author{W. T. Masselink}
\affiliation{%
Department of Physics, Humboldt-Universit\"{a}t zu Berlin, Newton-Str. 15, D-12489 Berlin, Germany
%\\This line break forced% with \\
}%

\author{F. Hatami}
\affiliation{%
Department of Physics, Humboldt-Universit\"{a}t zu Berlin, Newton-Str. 15, D-12489 Berlin, Germany
%\\This line break forced% with \\
}%

\date{\today}% It is always \today, today,
% but any date may be explicitly specified

\begin{abstract}
We describe the optical emission and the carrier dynamics of
an ensemble of self-assembled GaAs quantum dots embedded in GaP(001).
The QD formation is driven by the 3.6 $\%$ lattice mismatch between
GaAs and GaP in Stranski-Krastanow mode after deposition of more than 1.2
monolayers of GaAs. The quantum dots have an areal density between 6 and 7.6 $\times 10 ^{10}$ per cm$^{-2} $ and multimodal size distribution. The luminescence spectra show two peaks in the range of 1.7 and 2.1 eV. The samples with larger quantum dots have red emission and show less thermal quenching compared to the samples with smaller QDs. The large QDs luminescence up to room temperature. We attribute the high energy emission to  indirect carrier recombination in the thin quantum wells or small strained quantum dots, whereas the low energy red emission is due to the direct electron-hole recombination in the relaxed quantum dots.

\end{abstract}

\pacs{78.66.Fd, 78.67.Hc, 78.67.De, 78.60.Hk, 78.47.jd, 78.55.Cr}% PACS, the Physics and Astronomy
                             % Classification Scheme.
\keywords{III-V semiconductor, QDs, luminescence}%Use showkeys class option if keyword
                              %display desired
\maketitle

%*** introduction ***

Optical emitters are one of the key components of optoelectronics
and photonics. In spite of the rapid progress in light emitting
technology for the visible spectral range during the last decades,
the realization of efficient light emitters embedded in the silicon
environment remains a challenge. Several efforts have been focused
on monolithic integration of III-V light emitters with silicon.
Among the III-V compounds, GaP has the smallest lattice mismatch to
Si (less than 0.4\%). Conventionally, GaP is used in
manufacturing low-cost green to orange light-emitting diodes (LEDs). GaP is,
however, similar to Si, an indirect band gap semiconductor
($\text{E}_\text{g}^\text{ind} = 2.26$~eV) and useful light emission
from it is only possible by doping which results in low luminescence
efficiency. The challenge of high-efficiency light emitters based on GaP has
recently been approached by exploring the use of epitaxial quantum
structures embedded in this material. 
%Direct electron-hole recombination within a GaP matrix using 
%III-V nanostructures permit high quantum efficiencies.

The optical emission from several GaP-based QDs systems has been
investigated. The InP/GaP QD system with a lattice mismatch of 7.7\%
emits light at about 2~eV \cite{hatami01,hatami03} and light-emitting diodes based on InP/GaP quantum structures emitting in the red to green spectrum
have been demonstrated \cite{NT,hatami06}. However, it has been
found that the InP/GaP quantum structures do not always possess a direct
heterointerface and depending on the size and geometry of quantum
structures the heterointerface can be indirect in both \textbf{k}
and real space \cite{Goni03,Hatami2001}, or the lowest electron
states can be localized at the interface with a mixed $\Gamma-X$
character \cite{Zunger11}. The system InAs/GaP may provide a stable
direct heterointerface due to its higher conduction band offset
compared to the InP/GaP. However, the high lattice mismatch of about 11.4\% between InAs and GaP results not only in the formation of Volmer-Weber islands but also in dislocations \cite{Leon98}. Adding Ga into InAs reduces the large lattice mismatch and allows the growth of defect-free Stranski-Krastanow QDs. The (In,Ga)As/GaP QDs show photoluminescence (PL) in the red range \cite{song10,thanh11,stracke14} and, depending on the In concentration and the shape the dots, can have direct heterointerface \cite{robert12}. Light emitting diodes
and electrical injection QD laser have been demonstrated based on this materials system \cite{Lee,Heide14}. In the materials system (In,Ga)As/GaP, the GaAs/GaP has
the lowest lattice mismatch of 3.6\% but sufficient for formation of self-organized defect-free QDs in the Stranski$- $Krastanow mode \cite{nomura93}. Low-temperature PL of GaAs/GaP QDs at about 1.96~eV has been reported \cite{shamirzev10,abramkin12}. It has been shown that the relaxed GaAs/GaP QDs have a type-I heterointerface with the lowest electronic state at L minima of the GaAs conduction band \cite{shamirzev10}. The pseudomorphic GaAs/GaP quantum wells (QWs) and QDs, however, have type-II heterointerface \cite{abramkin12}.

In this letter, we demonstrate that by controlling the growth of GaAs/GaP QDs, the GaAs/GaP QDs can be changed from an indirect type-II or type-I system to a direct type-I system with light emission up to room temperature. Depending on the growth conditions, the optical emission of the structures peaks between 1.7 and 2.1~eV.

%*** Growth ***

Structures were grown in a Riber-32P gas-source molecular beam
epitaxy (GSMBE) system on GaP(001) substrates. After oxide desorption, a 0.5~$\mu$m GaP
buffer layer at a growth rate of 0.83 $ \mu$m/h was grown at 520$^\circ$C. The growth then was interrupted while the substrate temperature was reduced to
450$^\circ$C. Then, the desired thickness of GaAs was grown at a rate of 0.3 monolayers/s (ML/s). The resulting structures were capped by 50~nm of GaP. During growth, the process was monitored using reflection high-energy electron diffraction (RHEED). The surface of GaP showed a sharp (2$\times$4) reconstruction. The RHEED pattern remained streaky for the first ML of GaAs, indicating 2-dimensional growth and changed to the broken lines for thicker GaAs coverage, indicating the formation of QDs. Each sample contains one GaAs layer with a nominal coverage between 1.2 and 3.6 ML. (One ML of unstrained GaAs is 0.28~nm thick.) Samples for atomic force microscopy (AFM) were grown without the GaP cap layer.

%5.6533 $ \AA $

%%%%%%%%%%%%%%%%%%%%%%%%%%%%%%
\begin{figure}
\includegraphics[scale=0.32]{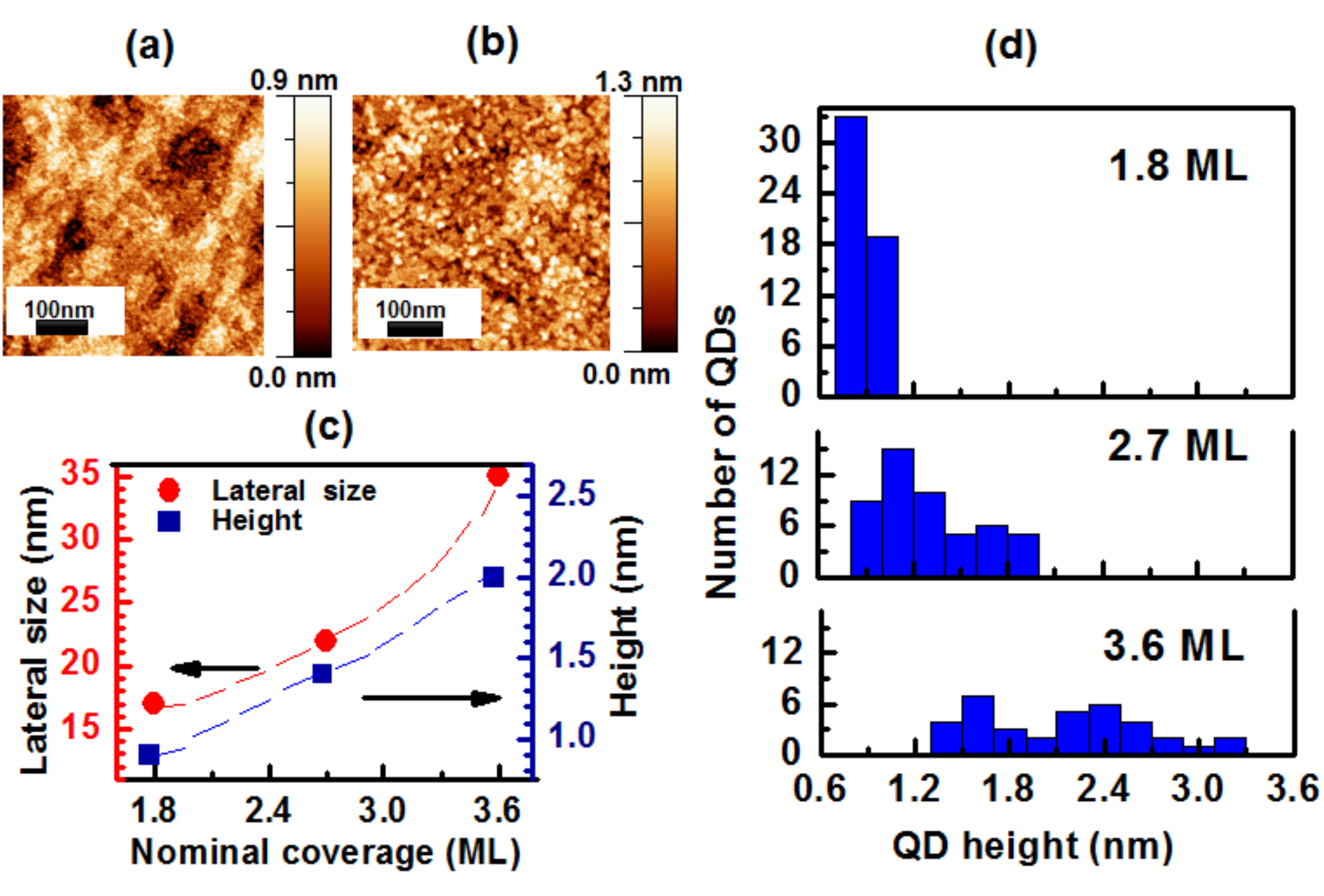}
\caption{} 
% (add the coverage to the pics????)
\end{figure}
%%%%%%%%%%%%%%%%%%%%%%%%%%%%%%

The density and size of QDs were characterized using a Bruker Icon Dimension AFM with peak force mode \cite{horcas07}. Fig. 1 depicts the AFM images of the samples with nominal coverage of 1.2 (Fig. 1a) and 1.8 MLs (Fig. 1b) of GaAs. According to the AFM
results, the formation of QDs occurs for GaAs nominal
thickness greater than 1.2 ML and confirms our RHEED observation. The lateral size and the average height of QDs increase with increasing nominal
thickness from 18 nm and 0.9 nm (1.8-ML GaAs) to 36 nm and 2 nm (3.6-ML GaAs). The data are summarized in Fig. 1c. The areal density of QDs  varies in the range of $6.0-7.6 \pm 0.2 \times 10 ^{10}$  cm$^{-2} $. The estimated thickness of the wetting layer using the AFM data is between 0.3 - 0.5 nm. The cross-section image of a capped 1.8-ML sample measured using high-angle annular dark-field scanning transmission electron microscopy (TEM) provides similar information about the wetting-layer thickness and the size of the QDs. Fig. 1d shows the distribution of the height of QDs for three different nominal coverage of GaAs between 1.8 to 3.6 ML. The distribution of both height and lateral size seems to be multimodal \cite{kissel00}. The similar size distribution has been reported for the GaAs/GaP QDs investigated by TEM \cite{shamirzev10}.

%%%%%%%%%%%%%%%%%%%%%%%%%%%%%%
\begin{figure}
\includegraphics[scale=0.15]{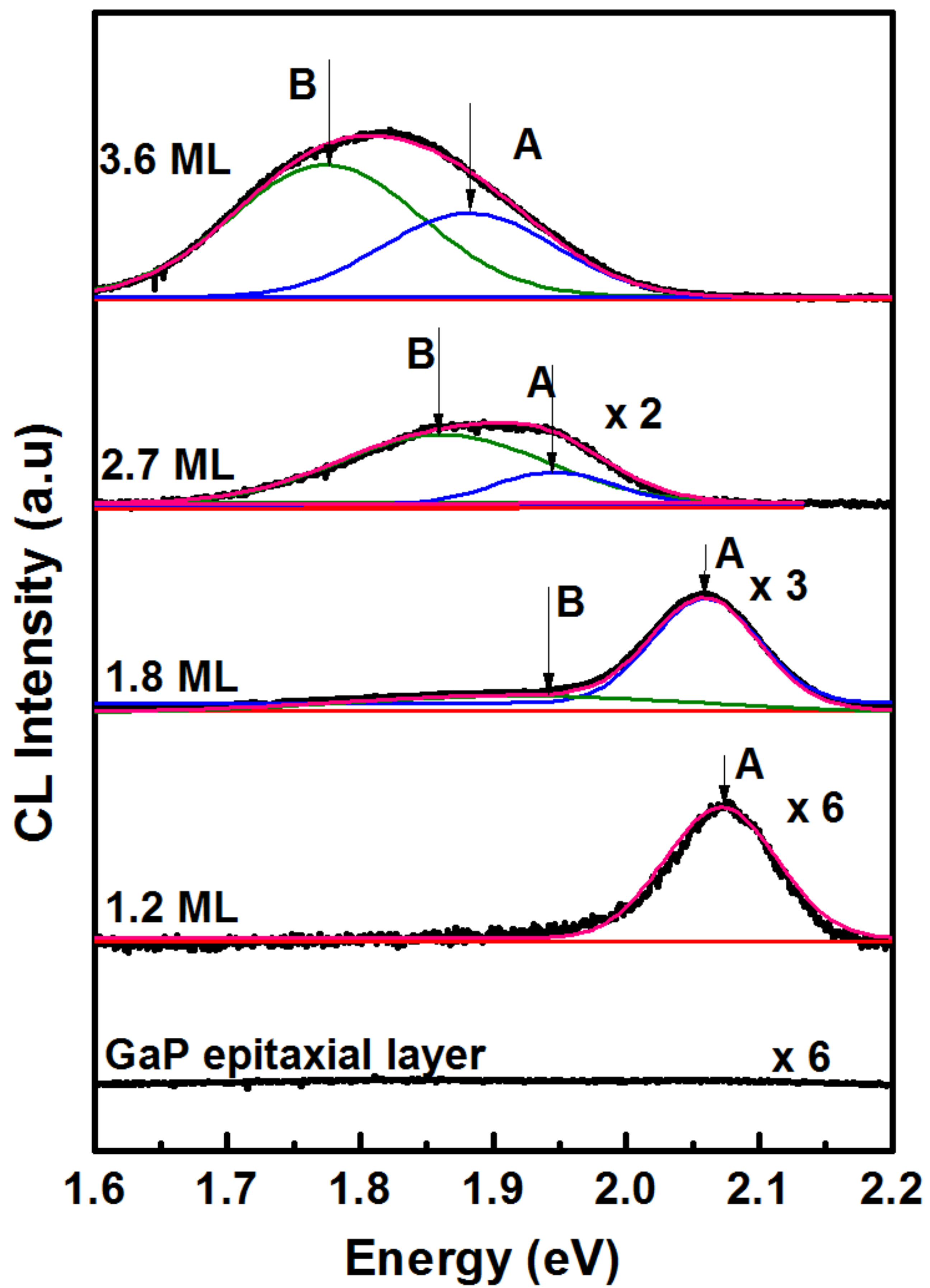}
\caption{}
\end{figure}
%%%%%%%%%%%%%%%%%%%%%%%%%%%%%%

Spectrally resolved cathodoluminescence (CL) measurements were carried out using a XiCLOne
(Gatan UK) CL system attached to an LEO 1530 (Carl-Zeiss) field-emission scanning electron microscope. The CL signal was detected with a Peltier-cooled CCD camera. Fig. 2 depicts the CL spectra (80 K) and their corresponding fits from four GaAs/GaP samples, each consisting of a single layer GaAs with nominal thickness between 1.2 and 3.6 MLs and a 50-nm GaP cap layer. In order to reduce the effect of the luminescence due to the GaP buffer, we applied an acceleration voltage of 5 kV that leads to an effective excitation depth of about 100 nm. The beam current was controlled by varying the microscope aperture size. The CL spectrum of a GaP epitaxial layer is also shown for comparison in Fig. 2. The CL spectrum of the 1.2-ML sample, which, according to the AFM results, consists of a single QW, demonstrates only one emission peak at 2.07 eV. Samples
with GaAs nominal thickness above 1.2 ML and containing QDs show, however, spectra that can be fitted using two Gaussian-shaped lines. For thicker GaAs nominal coverage, the total light output increases and the spectra have an overall redshift. The first emission peak at higher energies (position A) shifts from 2.06 (1.8-ML sample) to 1.88 eV (3.6-ML sample) and the second emission peak (position B) shifts
from 1.93 (1.8-ML sample) to 1.78 eV (3.6-ML sample).

Temperature dependent CL and PL measurements show that the 2.7-ML and 3.6-ML samples emit light up to room temperature, whereas the light output of the 1.2-ML and 1.8-ML samples become negligible for temperatures above 160~K. Apparently, the thermal quenching of the luminescence is less for the thicker GaAs coverage; e.g. in the temperature range between 10 and 300~K the light output of 2.7-ML sample decreases more than 30 times and for 3.6 ML sample only 6 times.

%%%%%%%%%%%%%%%%%%%%%%%%%%%%%%
\begin{figure}
\includegraphics[scale=0.20]{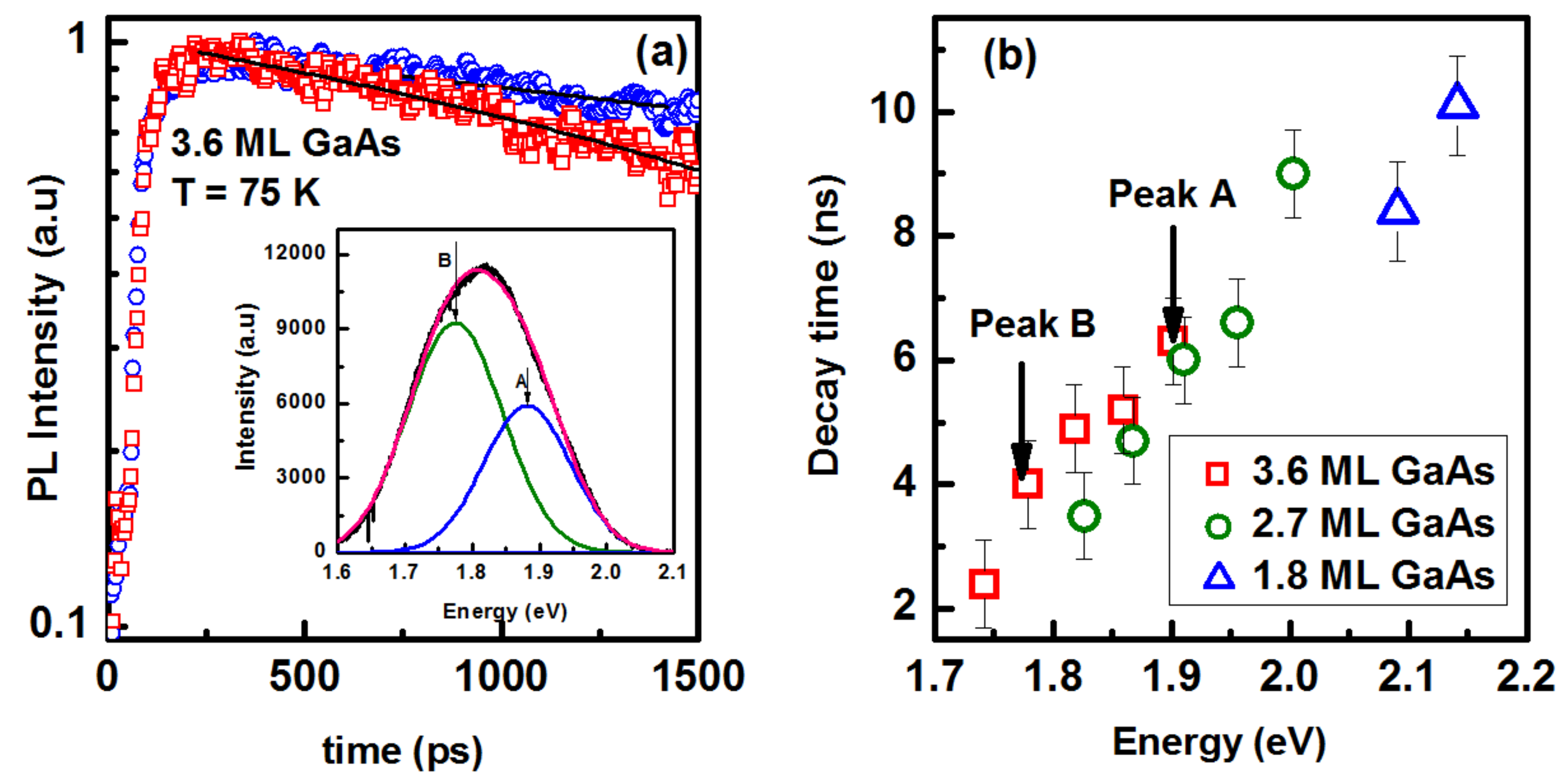}
\caption{ }
\end{figure}
%%%%%%%%%%%%%%%%%%%%%%%%%%%%%%
In order to investigate the carrier recombination dynamics and the origin of the luminescence, we carried out time-resolved PL between 5 and 300~K. The samples were excited by 80-fs pulses at 388 nm from a mode-locked laser producing an optical pulse train at 82 MHz. Fig. 3a depicts the PL intensity decay of the 3.6-ML sample at 75 K for two emission peaks, A and B. The transients were fitted by a simple exponential decay with a single time constant. Fig. 3b summarizes the energy dependence of the emission decay times for the three different samples represented in Fig. 2. The decay times for the peak A and B are highlighted. For all samples, higher emission energy is correlated with slower luminescence decay. As the nominal coverage of GaAs is increased from 1.8 ML to 3.6 ML, the decay time for the peak A decreases from 8.5 ns to 6 ns. The decay time at peak B changes from 3 ns to 2.4 ns. Increasing temperature results in the shorter decay time as a consequence of the thermally activated non-radiative decay processes; e.g. the 3.6-ML sample exhibits decay times at 300 K of about 3.1 and 0.9 ns for peaks A and B.

Our experimental results can be understood as follows: According to the structural investigations, the QW sample (1.2 ML) and the small QDs (less than 1 nm high) are fully strained. Due to the strain, the conduction band minimum of GaAs has a shift of 356 meV to higher energy and hence, the $\Gamma$-like electrons in the GaAs are located significantly higher than the X-valley in the GaP. In such a case, the favorable optical transition is indirect and occurs between the electrons (e) localized at the X state of GaP and the heavy-holes (hh) in GaAs. With increased the GaAs coverage, the QDs become larger and partially relaxed. Depending on the degree of relaxation, the bottom of the $\Gamma$-conduction band in GaAs shifts to lower energy. Finally for the fully relaxed QDs, the electrons at the $\Gamma$ state in GaAs are localized lower than the X-valley in the GaP matrix and the electron-hole transition becomes direct in both $\mathbf{k}$ and real space. Fig. 4a shows schematically the band alignment of the two extreme situations of a fully strained (type II and indirect) and a fully relaxed (type I and direct) GaAs/GaP heterostructure. The two arrows indicate the e-hh transition for both cases. From the AFM analysis, we know that for nominal thickness above 1.8 ML, a multimodal size distribution fits better to the investigated QDs than a single Gaussian distribution. We divide the QDs, therefore, depending on their height, into small, intermediate, and large QDs. Such arrangement fits to the lateral size distribution, too. The small QDs are strained with an indirect e-hh transition, whereas the large QDs are relaxed with a direct e-hh transition. The electrons localized in the QDs with intermediate size are expected to have a mixed $\Gamma-X$ nature. We consider that all of QDs contribute to the optical emission and they are the origin of peaks A and B in Fig. 2 for nominal thickness above 1.8 ML. Peak A is due to the smaller QDs and peak B to the larger QDs.

Our model is also experimentally supported by the behavior of the luminescence intensity for the samples with different GaAs coverage, in particular the ratio between peaks A and B. As can be seen in Fig. 2, an increased nominal coverage of GaAs is correlated with a larger total light output and more dominant peak B. We suggest that the increasing of the total light intensity is connected to a transition from a fully
strained indirect type-II QDs system (1.8 ML sample) to a relaxed system with direct type-I QDs (3.6-ML sample). For 2.7 and 3.6 ML samples, the large QDs are already relaxed and have higher intensity due to the stronger overlap between electrons and
holes wave functions in the type-I band alignment. The facts that both samples emit light up to room temperature and the faster decay time for peak B indicate a direct gap and type-I character of large QDs in the samples. The larger QDs have light emission at lower energy mainly due to the confinement effect. The small QDs, on the other hand, emit predomintly at position A with a slower decay time and faster thermal quenching, behaving like a type-II system. The connection between the increasing of size and the red shift in emission energy with the corresponding band alignment explain the decay time behavior presented in Fig. 3b. The probability to have a mixed direct and indirect transition, and finally a direct and type-I QD increases with increasing the size of the QDs. Larger dots with type-I band alignment have large overlap between the electron-hole wave functions and hence, faster decay time. The fastest measured decay time of about 2 ns is still higher than the reported values for the pure type-I systems, but it can be explained by the mixed direct-indirect nature of the optical transitions.

Generally, at the interface of two semiconductors, with their conduction band minima at two different valleys, the formation of localized interface states is expected. Such states may have also mixed $\Gamma-X$ character and lead to optical transition, particularly for a common anion heterointerface. However, in the case of GaAs/GaP with a common cation heterointerface the interface states are not localized in the gap and are not optically active, as predicted by atomic empirical pseudopotential calculations \cite{Zunger11}.

The appearance of peak A and B in luminescence spectra shows that both kinds of transition occur simultaneously in a mixed ensemble of type-I and type-II QDs. The excitation intensity dependence of the PL measurements in the range of 0.001 to 10 W/cm$^2$ verifies this picture and shows a remarkable 50 meV blueshift for peak A and a slight blueshift of about 6 meV for peak B for the 3.6-ML sample. The strong blueshift has already been reported for several type-II QDs \cite{Nakaema,ledentsov}.

% The density is 7e10, 7.6e10, and 6 e10 for 1.8, 2.7, and 3.6 ML.

%%%%%%%%%%%%%%%%%%%%%%%%%%%%%%
\begin{figure}
\includegraphics[scale=0.30]{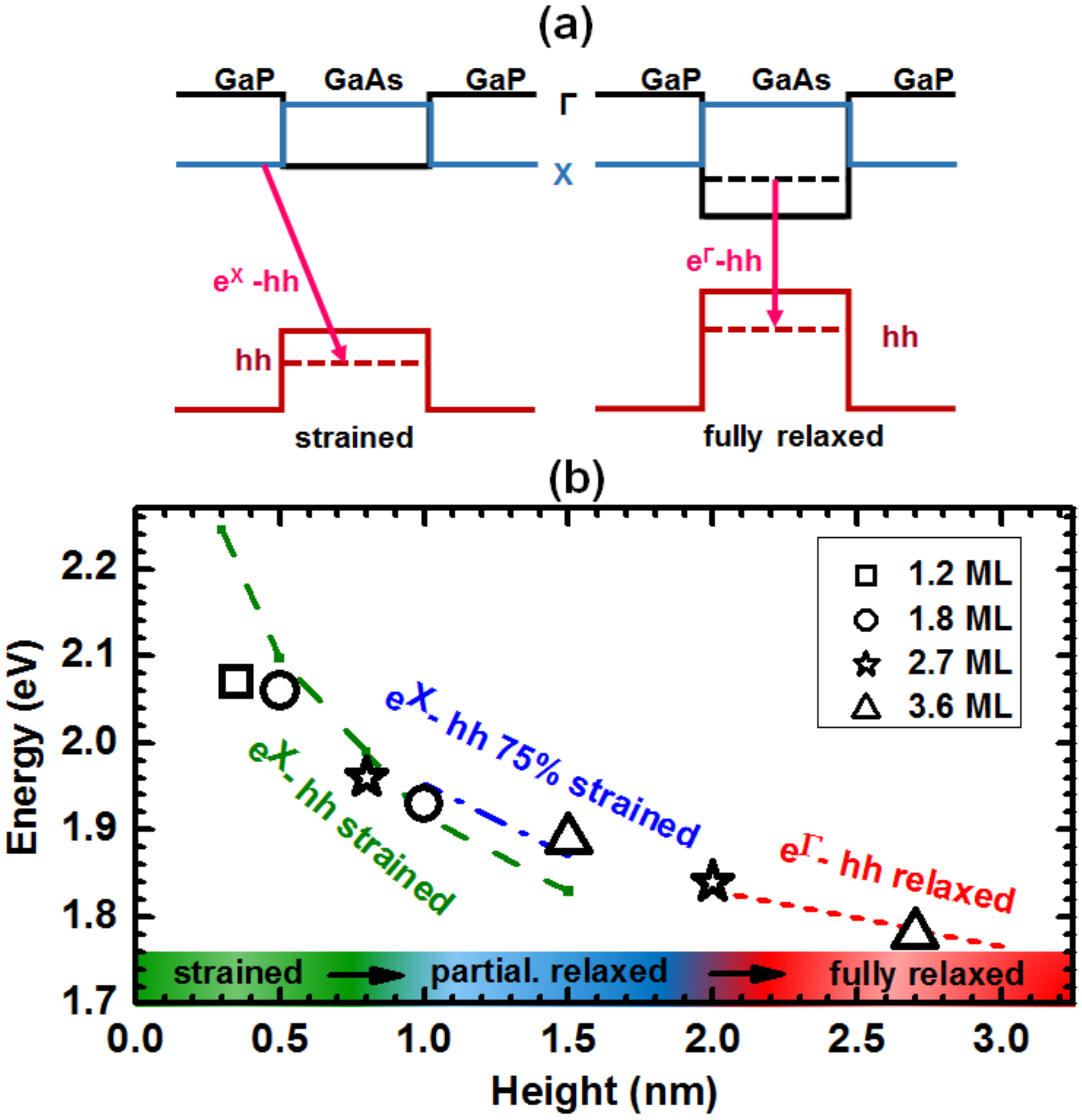}
\caption{  }
\end{figure}
%%%%%%%%%%%%%%%%%%%%%%%%%%%%%%
In order to corroborate our model, we calculated the energy states of
GaAs/GaP QWs and QDs  using the effective mass approximation for different relaxation degree  \cite{Birner07}. Since the height of the QDs is at least one order of magnitude smaller than their lateral size, the QDs were modeled as QWs with a thickness equivalent to the height of the QDs. According to the calculation, a direct e-hh transition in GaAs QDs is expected only if the QDs are higher than 2 nm and fully relaxed. Fig. 4b depicts the calculated energetically lowest and favorable e-hh transition for fully strained, 75\% strained, and fully relaxed QDs as a function of the height. The symbols show the peak positions of samples with nominal coverage between 1.2 ML to 3.6 ML. The corresponding height for every peak position was taken from AFM analysis and gives the average height of small and large QDs in every sample. The luminescence peak positions fit well to the calculated results.

%using the nexnanomate software

These results support our model for a transition from optically low-efficient indirect type-II strained QW and QDs to optically efficient direct type-I relaxed QDs
while the size of dots increases. The strain relaxation in the GaAs/GaP QDs occurs via the formation of edge dislocations, which do not act as nonradiative recombination centers and hence, allows the realization of high-efficient e-hh transition \cite{shamirzev10}. Similar characteristics have been also reported for the QDs based on the direct gap InGaAs embedded in the indirect GaP matrix \cite{stracke14}. 

Several Stranski-Krastanow QDs systems such as (In,Ga)As/GaAs and InAs/InP show luminescence spectra with two peaks, which are attributed to the QDs and to the wetting layer. However, the emission energy of GaAs/GaP QDs is very low for the strained wetting layer with an estimated thickness of $0.4\pm 0.1$ nm. Therefore, we believe that the origin of the light emission is the carrier recombination in the QDs. Only for 1.8-ML sample the origin of peak A might be the fully strained wetting layer.

In summary, we have demonstrated the optical emission of the GaAs/GaP QDs up to room temperature in the visible range. The luminescence spectra can be fitted using two Gaussians. The AFM analysis shows a multimodal size distribution for the QDs. The thin quantum wells and small dots are fully strained and have type-II band alignment with  indirect e-hh transition, whereas the large QDs are fully relaxed with type-I band alignment and direct e-hh transition.

%\section{Acknowledgments }
This research has been supported by the European Commission FP7-ICT-2013-613024-GRASP. S. Dadgostar thanks Yousef Jameel Foundation for the financial support.

%==========================================

\newpage

\section*{Figure captions}

Figure.1- AFM images of (a)
1.2 ML and (b) 1.8 ML GaAs/GaP. The scale bar is 100 nm. (c) The summary of the AFM analysis for the lateral size (circles and left axis) and height (squares and right axis) for different GaAs coverage. The dashed lines are
a guide to the eye. (d) Histograms
of height distribution for three different GaAs coverage.\\

Figure.2- CL spectra of GaAs/GaP quantum structures capped by 50 nm GaP
at 80 K with corresponding fits. CL spectrum of GaP buffer is also shown for
comparison. The measurements were performed using 5~keV excitation energy.\\

Figure.3- (a) Time-resolved PL spectra of the 3.6-ML sample for two emission peak maxima A (blue circles) and B (red squares) at 75~K with their corresponding fits (black lines). The two arrows in the inset show the energy position of peak A and B. The decay time of peak A is 5.5 ns and for peak B 2.4 ns.
(b) PL Decay times of 1.8, 2,7, and 3.6 ML samples for different emission energy at 75~K.\\

Figure.4- (a) Scheme of band alignment for a fully strained (type II and indirect) and a fully relaxed (type I and relaxed) GaAs/GaP heterostructure. The two arrows present the favorable e-hh transition. (b) The calculated energetically lowest e-hh transition for fully strained, 75\% strained, and fully relaxed QDs as a function of the height. The symbols show the luminescence peak positions of four samples. The height for every peak position was taken from AFM analysis and gives the average height of small and large dots for the QDs samples, and the thickness of the QW in 1.2-ML sample.

\end{document}